\documentclass[a4paper,12pt]{article}

\usepackage[T1]{fontenc}
\usepackage[utf8]{inputenc}
\usepackage[english]{babel}
\usepackage{lmodern}

\usepackage{authblk}
\usepackage{amsmath}
\usepackage{amssymb}
\usepackage{cite}
\usepackage{hyperref}
\usepackage{xcolor}
\usepackage{fancyhdr}

\renewcommand{\headheight}{28pt}

\title{Kira 1.2 Release Notes}
\author[,a]{Philipp Maierh\"ofer%
  \footnote{E-mail: \href{mailto:philipp.maierhoefer@physik.uni-freiburg.de}{%
                          philipp.maierhoefer@physik.uni-freiburg.de}}}
\author[,b]{Johann Usovitsch%
  \footnote{E-mail: \href{mailto:usovitsj@maths.tcd.ie}{usovitsj@maths.tcd.ie}}}
\affil[a]{Physikalisches Institut, Albert-Ludwigs-Universit\"at Freiburg,
          79104~Freiburg, Germany}
\affil[b]{Trinity College Dublin, School of Mathematics, Dublin 2, Ireland}
\date{}

\newcommand*{\kira}{\texttt{Kira}}

\newcommand*{\git}{\texttt{Git}}
\newcommand*{\gitlab}{\texttt{GitLab}}

\begin{document}

\fancypagestyle{firstpage}{\rhead{
  FR-PHENO-2018-15\\
  TCDMATH 18-18%
}}

\maketitle
\thispagestyle{firstpage}

\begin{abstract}
  We present the Feynman integral reduction program \kira{}\;\texttt{1.2} and
  describe its new features and other changes w.r.t.\ the previous versions.
  The main new features include a much faster equation generator, more flexible
  seed specification options, several predefined integral orderings, the
  reduction of systems of user-provided equations and a new technique to
  simplify coefficients by sampling variables over integers.
  Furthermore, we provide a collection of recommendations on how to use the
  program efficiently.
  This version has overall improvements in runtime for all reduction tasks
  compared to the previous versions of \kira{}.
\end{abstract}

\clearpage
\renewcommand{\headheight}{0pt}

\section{Introduction}

\kira{}, first introduced in \cite{Maierhoefer:2017hyi}, is an implementation of
the Laporta algorithm \cite{Laporta:2001dd} to reduce Feynman integrals to a
basis of master integrals.
In this approach, large systems of relations between Feynman integrals, namely
integration-by-parts \cite{Chetyrkin:1981qh} and Lorentz invariance
\cite{Gehrmann:1999as} identities, and symmetry relations, are generated and
solved by Gaussian elimination.
A complexity criterion defines an ordering on the Feynman integrals, so that
more complicated integrals are systematically expressed in terms of simpler
integrals.
Laporta's algorithm has become a standard method for the integral reduction task
in calculations of multi-loop scattering amplitudes e.g.\ for the Large Hadron
Collider, and constitutes one of the bottlenecks in such calculations.
Though alternative reduction techniques have been proposed and applied to
specific problems, see
e.g.\ \cite{Smirnov:2005ky,Lee:2013mka,Larsen:2015ped,Kosower:2018obg}, to date
programs based on Laporta's algorithm
\cite{Anastasiou:2004vj,vonManteuffel:2012np,Smirnov:2014hma} pose the only
general purpose tools suited for large scale applications.
The continuous improvement of these tools is thus essential to keep up with the
increasing demands in the applications.

In \kira{}, the solver for systems of linear equations is assisted by modular
arithmetic to remove redundant equations from the system and to select
subsystems of equations in order to reduce specific selected integrals.
The employed forward elimination tries to minimise the number of elimination
steps, and the simplification of coefficients is optimised e.g.\ by recursive
pairwise combinations of coefficients.
The details of the algorithms used are described in \cite{Maierhoefer:2017hyi}.

In section \ref{sect:installation} of this article we describe how to obtain
\kira{} and how to compile and install it on your system.
Section \ref{sect:news} explains new features, fixes and changed default
behaviour in \kira{}\;\texttt{1.2}.
In section \ref{sect:bestpractices} we provide a collection of tips on how to
use \kira{} efficiently.
Appendices \ref{gettingmeson} and \ref{compiledependencies} provide further help
regarding the installation of \kira{} and its dependencies.

Note that this is not a complete user manual.
In particular, for the conventions used in the topology and kinematics
definition, please refer to the original \kira{} publication
\cite{Maierhoefer:2017hyi}.
Even though no systematic introduction on the structure of job files is provided
in these release notes, together with the examples that are shipped with the
source code, they should contain enough information to cover the majority of use
cases.

\clearpage

\section{Installation}
\label{sect:installation}

\kira{} can either be built using the \texttt{Autotools} build system or the
\texttt{Meson} build system \cite{Meson}.
We recommend \texttt{Meson} if a sufficiently recent version is available on
your system.

\subsection{Prerequisites}

\noindent
\textit{Platform requirements}\smallskip\\
Linux x86\_64 or Mac OS X.

\medskip
\noindent
\textit{Compiler requirements}\smallskip\\
A \texttt{C++} compiler supporting the \texttt{C++14} standard and a \texttt{C}
compiler supporting the \texttt{C11} standard, e.g.\ \texttt{gcc} version 5.1 or
later, or \texttt{LLVM} version 3.4 or later.

\medskip
\noindent
\textit{Dependencies}\smallskip\\
\kira{} requires the following packages to be installed on the system:
\begin{itemize}
  \item \texttt{GiNaC} \cite{Bauer:2000cp,Vollinga:2005pk} (which itself
    requires \texttt{CLN} \cite{CLN}),
  \item \texttt{yaml-cpp} \cite{YAMLCPP},
  \item \texttt{zlib} \cite{ZLIB}.
\end{itemize}
In addition, the program \texttt{Fermat} \cite{Fermat} is required to run
\kira{}.
If the \texttt{Fermat} executable is not found automatically by \kira{} or a
specific \texttt{Fermat} installation should be used, the path to the
\texttt{Fermat} executable can be provided by setting the environment variable
\texttt{FERMATPATH=/path/to/fer64}.

Please note that \texttt{GiNaC}, \texttt{CLN} and \texttt{yaml-cpp} must have
been compiled with the same compiler which is used to compile \kira{}.
Otherwise the linking step will most likely fail.
If you are using the system compiler, you can usually install these packages via
your system's package manager.
However, if you are using a different compiler, this means in practice that you
also have to build these packages from source and set the environment
variables \texttt{C\_PATH}, \texttt{LD\_LIBRARY\_PATH} and
\texttt{PKG\_CONFIG\_PATH} according to the installation prefix.
See appendix \ref{compiledependencies} for more details.

\subsection{Obtaining Kira}

The latest version of \kira{} is available from our \git{} repository at
\gitlab{} under the URL \url{https://gitlab.com/kira-pyred/kira}.
To obtain the source code of the latest release version, clone the repository
with
\begin{verbatim}
  git clone https://gitlab.com/kira-pyred/kira.git -b release
\end{verbatim}
checking out the release branch.
Release versions are also available as \git{} tags (starting with
\kira{}\;\texttt{1.2}).
The master branch of the repository contains the latest pre-release version,
receiving more frequent updates with new features and fixes.
To obtain the source code of the latest pre-release version, clone the
repository with
\begin{verbatim}
  git clone https://gitlab.com/kira-pyred/kira.git
\end{verbatim}
checking out the master branch.
Please refer to the changelog for information about what changed w.r.t.\ the
release.
Packages of the release versions as compressed \texttt{Tar} archives are
available from\\[1.2ex]
\hspace*{2.5ex}\url{https://www.physik.hu-berlin.de/de/pep/tools}.

\subsection{Compiling Kira with the Autotools build system}
\label{installwithautotools}

If you obtained \kira{} from the \git{} repository, you first need
to run
\begin{verbatim}
  autoreconf -i
\end{verbatim}
Then compile and install with
\begin{verbatim}
  ./configure --prefix=/install/path
  make
  make install
\end{verbatim}
where the optional \texttt{-{}-prefix} argument sets the installation prefix.

\subsection{Compiling Kira with the Meson build system}

Since \kira{}\;\texttt{1.2} one may optionally use the \texttt{Meson} build
system instead of \texttt{Autotools} to build \kira{}.
\begin{verbatim}
  meson --prefix=/install/path builddir
  cd builddir
  ninja
  ninja install
\end{verbatim}
where \texttt{builddir} is the build directory.
Specifying the installation prefix with \texttt{-{}-prefix} is optional.
\texttt{Meson} version 0.46 or later is recommended\footnote{Some older versions
may work as well, but this has not been tested.}.
See appendix \ref{gettingmeson} for more information about how to obtain a
current version of \texttt{Meson}.

\section{Changes and new features in Kira 1.2}
\label{sect:news}

\subsection{Build system}

\begin{itemize}
  \item A new build system based on \texttt{Meson} has been introduced.
  \item The dependency on \texttt{OpenMP} has been removed.
        Instead, \kira{} now uses threads directly.
  \item Support for Mac OS X has been improved. \kira{} now compiles with
        Apple's version of \texttt{LLVM} (which lacks \texttt{OpenMP} support).
\end{itemize}

\subsection{New equation generator}

The generator for integration-by-parts, Lorentz invariance and symmetry
equations has been rewritten.
The performance of the equation generator has been improved by typically
three orders of magnitude.

\subsection{Changed default options}
\label{sect:defaultoptions}

\begin{itemize}
  \item The option \texttt{data\_file:\;false} in the job file is now default,
        i.e.\ the reduction rules are no more written to plain text files.
        This reduces disk usage by approximately a factor $2$.
        The old behaviour can be restored by setting \texttt{data\_file:\;true}.
  \item The option \texttt{conditional:\;true} in the job file is now default.
        With this option (introduced in \kira{}\;\texttt{1.1}), aborted
        reductions are automatically resumed when the job is restarted.
  \item The command line option \texttt{-{}-algebra} has been removed.
        This option was used to choose between two different algorithms to
        simplify a sum of terms.
        The choice is now made automatically for every simplification based on
        the number of terms in the sum and the size of the terms.
\end{itemize}

\subsection{New seed notation}
\label{seedselection}

A more flexible notation for seeds has been introduced.
\begin{verbatim}
  jobs:
    - reduce_sectors:
        reduce:
          - {topologies: [T1,...], sectors: [S1,...],
             r: rmax, s: smax, d: dmax}
          - {...} # can be used multiple times
\end{verbatim}
\begin{itemize}
  \item \texttt{rmax}: the maximal sum of positive propagator powers in the
        seed.
  \item \texttt{smax}: the maximal negative sum of negative propagator powers
        in the seed.
  \item \texttt{dmax} (optional): the maximal number of dots.
        On a sector with $L$ lines ($L$ is the number of propagators with
        positive integer exponents), seeds will be generated up to
        $r=\min(\texttt{rmax},L+\texttt{dmax})$.
        If omitted, only \texttt{rmax} will limit $r$.
        This option should usually only be used in integral selectors, but not
        in IBP seeds.
  \item \texttt{topologies} (optional): \texttt{T1} is the name of a topology.
        Several topologies may be selected.
        If omitted, all topologies defined in the project will be selected.
  \item \texttt{sectors} (optional): \texttt{S1} is a sector number.
        Several sectors may be selected.
        If omitted, the sectors specified as \texttt{top\_level\_sectors} in the
        topology definition in \texttt{integralfamilies.yaml} will be used
        (individually for each topology).
\end{itemize}
The same notation can be used in \texttt{select\_mandatory\_recursively} and the
export options \texttt{kira2form} and \texttt{kira2math}.
\medskip\\
\noindent
\textit{N.B.:} The old seed notation may still be used.
However, we recommend using the new notation.
See the section about best practices for the advantages of the new notation.

\subsection{Integral ordering}
\label{sect:integralordering}

The integral ordering can now be chosen by the user from eight predefined
orderings. The ordering can either be chosen on the command line with the option
\texttt{-{}-integral\_ordering=<value>} or in the job file with the option
\texttt{integral\_ordering:\;<value>}.

See \texttt{examples/topo4/advanced\_jobs\_reduction.yaml} for an example.
Previous versions of \kira{} used ordering 1 exclusively, which is still the
default.
\medskip\\
\noindent
\textit{Integral orderings}
\begin{enumerate}
  \item (default) Scalar products are regarded as simpler as dots.
        If there is more than one master integral in a sector, this
        produces a basis with irreducible scalar products (negative indices)
        and no increased propagator powers (positive indices greater than 1).
        If master integrals with increased propagator powers appear this usually
        indicates that the number of scalar products in the seed was chosen too
        small.
  \item Dots are regarded as simpler than scalar products.
        This produces a basis with only positive indices. 
        If master integrals with irreducible scalar products appear this usually
        indicates that the value of \texttt{rmax} in the seed was chosen too
        small.
  \item The first complexity criterion is the sum of dots and scalar products.
        If the sum is equal, scalar products are regarded as simpler.
  \item Like 3., but if the sum is equal, dots are regarded as simpler.
\end{enumerate}
Furthermore, in orderings 1--4, sectors are ordered w.r.t.\ the sector number.
Orderings 5--8 are the same as 1--4, but with sectors ordered by the number of
lines first.
Furthermore, in orderings 5--8, subsectors of the defined top-level sectors are
always regarded as simpler than sectors outside the top-level sectors with the
same number of lines.
This prevents a possible issue which is described in section
\ref{sect:magicrelations}.

Note that changing the integral ordering may affect the runtime significantly in
either direction, depending on the topology and its exact parametrisation.
We recommend to investigate the behavior in a test run.

\subsection{Basis change}

The feature \texttt{preferred\_masters} (introduced in \kira{}\;\texttt{1.1}) to
provide a list of integrals which should be preferred as master integrals has
been improved and bugs have been fixed:
\begin{itemize}
  \item Eliminating redundant integrals from the list of preferred master
        integrals now works.
  \item In some cases, the mapping of integrals was done incorrectly and an
        integral with all indices zero appeared as a master integral.
\end{itemize}
Note that a posteriori basis changes are not supported, i.e.\ preferred masters
must be given before the reduction is performed.

\medskip
\noindent
\textit{N.B.:} The option \texttt{select\_masters} is an alias of
\texttt{preferred\_masters}.

\subsection{Topology initialisation}

In the first run, all topologies which are defined in
\texttt{integralfamilies.yaml} will be initialised.
I.e.\ trivial sectors, symmetries and mappings will be determined for all
topologies in the order in which they are defined.
If an individual topology is reduced, master integrals will now be mapped to
previous topologies, even if they are not reduced in the same run.

In previous \kira{} versions, only topologies selected in the job file were
initialised (in the order they were given in the job file).
This lead to an error if, in a subsequent run, a topology was added to the job
file.
Furthermore, the mapping to previous topologies was only done when the
topologies were reduced together.

\subsection{Improved symmetry detection}

\begin{itemize}
  \item The detection of symmetries due to the permutation of external momenta
        now takes into account sign flips of the momenta.
        Only symmetries that leave the invariants unchanged are considered.
  \item Previous versions of \kira{} discarded certain symmetries if they were
        expected to be redundant.
        This has been disabled, because in some cases symmetries which are not
        redundant were dropped.
        Now all symmetries are always taken into account.
  \item In certain cases, \kira{}\;\texttt{1.1} tried to use symmetries which do
        not permute the propagators but shift their masses.
        While the symmetries are in principle valid, they are not handled
        correctly.
        Since these symmetries are unwanted in a reduction, they have been
disabled.
\end{itemize}

\subsection{Export of reduction rules to Mathematica or Form}

In \kira{}\;\texttt{1.1}, reduction rules for integrals in trivial sectors were
missing in the exported file. This has been fixed.

\subsection{Sectorwise forward elimination}

For large systems it may be helpful to perform the forward elimination
sectorwise.
To activate this, set \texttt{run\_triangular:\;sectorwise} (instead of
\texttt{run\_triangular:\;true}) in your job file.
If \texttt{run\_triangular:\;sectorwise} is set and a run is aborted before the
triangular form is achieved, \kira{} will resume the forward elimination when
the job is restarted.

\subsection{Algebraic reconstruction}

In reduction problems where the coefficients contain at least two variables, the
new option \texttt{algebraic\_reconstruction:\;true} can be used to enable the
algebraic reconstruction of coefficients sampled over integers
\cite{Larsen:2015ped,Boehm:2017wjc,Boehm:2018fpv}.
The algorithm is only applied during the back substitution.

If the option \texttt{algebraic\_reconstruction:\;true} is set \kira{} will
decide to which coefficients the algorithm will be applied based on the number
of terms in the expression and the size of the expressions.

\noindent
For an example see \texttt{examples/topo4/advanced\_jobs\_reduction.yaml}.

\subsection{Solving user-defined systems of equations}

One of the most requested features by \kira{} users is the possibility to use
the equation solver on externally provided equations%
\footnote{Employing some undocumented tricks this has always been possible and
in fact been done by several users, e.g.\ in \cite{Harlander:2018zpi}.}.
In \kira{}\;\texttt{1.2} an interface is provided to solve systems of
user-provided linear equations.
The option \texttt{reduce\_user\_defined\_system} comes in two flavours:
\begin{itemize}
  \item Import the user-defined system of equations without preparing the config
    files \texttt{integralfamilies.yaml} and \texttt{kinematics.yaml}.
    The system will be sorted according to \kira{}'s equation ordering and then
    be solved.
    Topologies will be automatically defined with the names used in the equation
    file in the order of their appearance.
    For integrals with more than one integer index, the orderings defined in
    section \ref{sect:integralordering} can be used.
  \item Prepare the config files in the \texttt{config} folder in the same way
    as you would in an ordinary integration-by-parts reduction.
    Based on the definitions in the config files, trivial integrals will be set
    to zero in the imported system.
    The system will be sorted and then solved.
\end{itemize}
Options like \texttt{select\_integrals}, \texttt{preferred\_masters} and those
to control the individual reduction steps apply as usual.

In cases where the objects in the equations are not integer-indexed Feynman
integrals or where a different ordering is desired than \kira{}'s build-in
integral orderings, single-indexed objects may be used, e.g.\ \texttt{T[n]},
where \texttt{n} is a 32-bit signed integer. In the default ordering,
\texttt{T[0]} will be regarded as the simplest object, followed by descending
negative indices. Positive indices are regarded as more complex than any
negative index with increased complexity for ascending values. This way,
$2^{32}$ distinguishable objects may be defined per ``topology''.

Some examples can be found in \texttt{examples/userDefinedSystem1} and
\texttt{examples/userDefinedSystem2}.

Possible applications include but are not limited to integrals appearing in
amplitudes that do not have the standard form of Feynman integrals, or
coefficients in the expansion of Feynman integrals in some parameter.

\subsection{Finding ``magic relations''}
\label{sect:magicrelations}

Some relations between Feynman integrals are only found when sectors with more
lines than the integrals themselves are taken into account.
In some cases, those higher sectors may not appear in the physical problem at
hand.
A straight-forward way to find these relations is, of course, to include those
higher sectors in the reduction (possibly with a lower seed) and excluding them
in the integral selection.

However, if the option \texttt{top\_level\_sectors} is used in the topology
definition, the symmetrisation of higher sectors will be skipped.
The new option \texttt{magic\_relations:\;true} in the topology definition
forces \kira{} to include symmetries of sectors which are not subsectors of the
top-level sectors in a way that the sectors are preferably mapped onto the
top-level sectors.
From experience it is sufficient to symmetrise only sectors with up to the same
number of lines as the top-level sectors.

An example is provided in \texttt{examples/topbox} where the number of master
integrals in sector $93$ is reduced from $5$ to $4$ when sector $127$ is
included in the seed and the option \texttt{magic\_relations:\;true} is set.

Note that when top-level sectors are embedded into a topology which do not have
a sector number expressible as $2^n-1$ with integer $n$, it may happen that such
``magic relations'' reduce a selected integral to integrals which lie outside of
the top-level sectors. In such cases, the number of master integrals may
increase, because these additional sectors will then be reduced as well. This
can be prevented by using a different sector ordering which regards top-level
sectors and their subsectors as simpler than all other sectors. The integral
orderings $5$--$9$ as described in section \ref{sect:integralordering} exhibit
this property and may therefore resolve the issue.

\subsection{Merging of database files}
\label{sect:merge}

\kira{} now provides the job file option \texttt{merge} to merge several
reduction databases into a single database file.

If a reduction was done independently to different subsets of master integrals
with the option \texttt{select\_masters\_reduction}, this can be used to merge
the database files with the complementary parts of the reduction into one
database with the complete reduction.
For an example see section \ref{subsec:reduct2individual} and the example
\texttt{examples/topo7\_parallel}.
In this example, we perform the reduction to the first half of the master
integrals in the project directory \texttt{examples/topo7\_parallel} and to the
second half in the directory \texttt{examples/topo7\_parallel2}.
Then we can merge the database from \texttt{topo7\_parallel2} into the one from
\texttt{topo7\_parallel} by running the job file
\begin{verbatim}
  jobs:
   - merge:
      files2merge:
       - ../topo7_parallel2/results/kira.db
\end{verbatim}
in the directory \texttt{examples/topo7\_parallel}.
Note that the path to the database to merge into is not listed in this file, but
implicitly taken from the project directory from which the merge job is
executed.
Afterwards, the fully reduced integrals can be extracted from the merged
database with \texttt{kira2math} or \texttt{kira2form} as usual.

\subsection{Cut propagators}

For completeness, we describe the option \texttt{cut\_propagators} already
introduced in \kira{}\;\texttt{1.1}.
To declare specific propagators of a topology as ``cut'', set the option
\begin{verbatim}
  cut_propagators: [n1,n2,n3,...]
\end{verbatim}
for the topology in \texttt{integralfamilies.yaml}, where
\texttt{[n1,n2,n3,...]} is the list of the numbers of the cut propagators
(propagators are numbered starting with 1). Example:
\begin{verbatim}
  integralfamilies:
    - name: "topo7"
      loop_momenta: [k1,k2]
      top_level_sectors: [127]
      propagators:
        - ["-k1", 0]         #1
        - ["k2", 0]          #2
        - ["-k1+k2", 0]      #3
        - ["k1+q2", "m2^2"]  #4
        - ["k2-p2", 0]       #5
        - ["-k1+p1+p2", 0]   #6
        - ["k2-p1-p2", 0]    #7
        - ["k1-p2", 0]       #8
        - ["k2-q2", 0]       #9
      cut_propagators: [3,4]
\end{verbatim}
Here, the 3rd and 4th propagator will be treated as cut. This means that during
the reduction all integrals in which a cut propagator has non-positive power are
set to zero.

\section{Best practices}
\label{sect:bestpractices}

\subsection{Topology definition}

If a topology has only one top-level sector, order the propagators such that the
sector number is $2^n-1$ (if the sector has $n$ lines), i.e.\ the propagators at
the end of the list appear only as irreducible scalar products.
Define the propagators in such a way that
\begin{enumerate}
  \item propagators have short linear combinations of loop momenta and the
        shortest linear combinations appear earlier in the list of propagators,
  \item analogously, keep linear combinations of external momenta short with
        ascending length,
  \item massless propagators appear earlier in the list.
\end{enumerate}
Always define \texttt{top\_level\_sectors} for the topologies.
This will ensure that sectors are only mapped on subsectors of the defined
top-level sectors.
Also, these are the sectors which are automatically selected when omitted in the
seed specification (see section \ref{seedselection}).
Sectors which are not subsectors of the defined top-level sectors will not be
symmetrised or mapped.
This reduces the amount of time spent on symmetry detection.
Note that if you are trying to find relations by reducing sectors which are
higher than those in the physical problem, the higher sectors must be included
in the top-level sectors in order to symmetrise them.
See also ``magic relations'' in section \ref{sect:magicrelations}.

These are only rough guidelines based on experience.
Finding the optimal topology representation is a non-trivial task and we do not
know a general prescription.
If you are dealing with a level of complexity where the topology representation
may be crucial to complete the reduction successfully on the available hardware,
try different representations and run smaller jobs (i.e.\ with smaller seeds)
and measure the effect.

\subsection{Integral selection}

Always use the options to select the integrals that should be reduced:\\
\texttt{select\_mandatory\_list} and/or \texttt{select\_mandatory\_recursively}
(see e.g.\ \texttt{examples/topo7/jobs2.yaml}).
\medskip\\

\noindent
\textit{Recommendation:}
\smallskip\\
Provide a list of all integrals needed for the amplitude and (if applicable) the
differential equations and pass it to \texttt{select\_mandatory\_list}.
If you intend to (manually) perform a transformation of the integral basis after
the reduction, use the option \texttt{select\_mandatory\_recursively} to select
integrals up to a certain number of dots and scalar products that you expect to
suffice for the transformation.

\medskip
When mandatory selectors have been defined, \kira{} prints the list of master
integrals after the selection has been performed and before starting the forward
elimination.
This step is much faster than the full reduction, i.e.\ one can obtain the list
of master integrals in a quick run.
It is possible to interrupt the reduction after printing the master integrals
by setting the option \texttt{run\_initiate:\;true} in the job file
and (if present) \texttt{run\_triangular} and \texttt{run\_back\_substitution}
to \texttt{false} in order to stop.

In time-consuming reductions, we recommend to always check that this list of
master integrals looks sensible before running the full reduction, e.g.\ that it
does not contain unexpected integrals like integrals that lie on the seed edge
although there are no masters in the same sector with one dot or one scalar
product less.

The reported list of master integrals at this stage contains all integrals which
appear in the reduction formulas of mandatory selected integrals and the
selected integrals themselves if there was no reduction formula found.
Of course, these master integrals can be used to determine integrals which enter
the differential equations for the master integrals before doing the actual
reduction.

Optional selectors as described in \cite{Maierhoefer:2017hyi} are not available
in \kira{}\;\texttt{1.2}.
They may be reintroduced in a later release.

\subsection{Advanced seed selection}

With the new seed selector described in section \ref{seedselection}, one may
generate and select a system of equations where the subsectors may have more
dots or scalar products than the higher sectors (with more lines).

\medskip
\noindent
See \texttt{example/topo7/jobs2.yaml}:
\begin{verbatim}
jobs:
 - reduce_sectors:
    reduce:
     - {topologies: [topo7], sectors: [127], r: 7, s: 2}
     - {topologies: [topo7], sectors: [63], r: 8, s: 3}
    select_integrals:
     select_mandatory_recursively:
      - {topologies: [topo7], sectors: [127], r: 7, s: 2, d: 0}
      - {topologies: [topo7], sectors: [63], r: 8, s: 3, d: 2}
\end{verbatim}
Here we demonstrate that we can generate a system of equations for the sector 63
with up to \texttt{r=8} and 3 scalar products while the higher sector 127 is
reduced up to \texttt{r=0} and 2 scalar products.
Such a reduction is impossible with the old seed selector notation.
Also the job files are shorter and topologies and sectors are chosen
automatically according to the topology definition.

\medskip
\noindent
See \texttt{example/topo7/jobs1.yaml}:
\begin{verbatim}
  jobs:
   - reduce_sectors:
      reduce:
       - {r: 7, s: 2}
\end{verbatim}
where \texttt{topo7} is implicitly chosen with sector 127 and \texttt{topo7x}
with sector 508 as defined in the topology definition with
\texttt{top\_level\_sectors} set.
An equivalent job file with the old seed notation looks like this:
\begin{verbatim}
  jobs:
   - reduce_sectors:
      sector_selection:
       select_recursively:
        - [topo7,127]
        - [topo7x,508]
      identities:
       ibp:
        - {r: [t,7], s: [0,2]}
\end{verbatim}

\subsection{Resuming aborted reduction jobs}
\label{sect:checkpoints}

In the following stages of the reduction \kira{} automatically creates
checkpoints from which an aborted calculation can be resumed:
\begin{itemize}
  \item After the search for the symmetries and trivial sectors of every
    topology.
  \item After generating the system of equations and selecting a subsystem
    of linearly independent equations.
  \item After the forward elimination, i.e.\ when the system is in triangular
    form.
    In case the option \texttt{run\_triangular:\;sectorwise} has been set, the
    state will be saved after every completed sector.
    Usually, the forward elimination is quite fast.
    If checkpointing is required anyway, \texttt{run\_triangular:\;sectorwise}
    must be set.
  \item During the back substitution, intermediate results will be committed to
    the database regularly.
\end{itemize}
When an aborted job is restarted, it will automatically continue from the last
checkpoint if the option \texttt{conditional:\;true} is set.
This is the default unless the option \texttt{select\_masters\_reduction} is
used (see section \ref{subsec:reduct2individual} for details).
Symmetries and trivial sectors will always be reused, independent of
\texttt{conditional}.

With the following options, different reduction steps may be started
individually:
\begin{itemize}
  \item \texttt{run\_symmetries:\;true/false},
  \item \texttt{run\_initiate:\;true/false},
  \item \texttt{run\_triangular:\;true/false/sectorwise},
  \item \texttt{run\_back\_substitution:\;true/false}.
\end{itemize}
Note that the bulk of the CPU time usually goes into the back substitution.
However, there is currently no option to terminate a job gracefully after a
specified amount of time.
Instead the job may just be terminated by the means of the operating system
(e.g.\ by a batch system with a time limit per job).

\subsection{Individual reductions to selected master integrals}
\label{subsec:reduct2individual}

Since \kira{}\;\texttt{1.1} it is possible to split the back substitution into
several reduction jobs where each job calculates the coefficients of a given
subset of master integrals only, effectively setting all other master integrals
to zero.
The full reduction is then obtained by merging the results from the individual
reductions (i.e.\ summing the right-hand sides of the reduction formulas for
each reduced integral).
Similar approaches have been described in \cite{Boehm:2018fpv,Chawdhry:2018awn}.
There are two scenarios in which this technique is useful.
\begin{itemize}
  \item Reduce the memory consumption during the back substitution by
    calculating coefficients for different master integrals sequentially in a
    single run.
    In order to use this feature, create an entry of the following form in your
    job file:
\begin{verbatim}
select_integrals:
 select_masters_reduction:
  - [topo7, [1,3,5,7,9,11,13,15,17,19,21,23,25,27,29,31]]
  - [topo7, [2,4,6,8,10,12,14,16,18,20,22,24,26,28,30]]
\end{verbatim}
    In this example, we have $31$ master integrals for the example topology
    \texttt{topo7} and the reduction will be performed in two sequential steps,
    first to the master integrals with odd numbers, then to those with even
    numbers.
    The merging into complete reduction formulas will be done automatically in
    this case.
    Note that in this mode resuming a reduction job with
    \texttt{conditional:\;true} is not possible. If resuming is forced
    nevertheless, it will most likely produce wrong results.
  \item Parallelise the reduction across different machines by letting each
    machine handle the coefficients of a subset of the master integrals.
    In order to do this, some manual work is required.
    We recommend to run the forward elimination (option
    \texttt{run\_triangular:\;true} or \texttt{sectorwise}) on one machine and
    then create a copy of the working directory for each back substitution job.
    Then run the back substitution jobs on all machines in parallel.

    In this mode resuming a reduction is possible (for every back substitution
    job individually). Note that if \texttt{select\_masters\_reduction} is used,
    the option \texttt{conditional:\;true} must be set explicitly to resume a
    job (checkpoints are created in any case, see section
    \ref{sect:checkpoints}).
\end{itemize}
It lies in the user's responsibility to make sure that the combined sets of
master integrals listed in \texttt{select\_masters\_reduction} is complete.
This means that the list of master integrals must be determined before
performing the back substitution.
The fastest way to find the master integrals with \kira{} is to run a job with
the option \texttt{run\_initiate:\;true} and perform an integral selection with
the options \texttt{select\_mandatory\_recursively} and/or
\texttt{select\_mandatory\_list}.
The list of master integrals will then be written to the file \texttt{master} in
the directory \texttt{results/TOPO} for topology \texttt{TOPO}.
The numbers of the master integrals in the option
\texttt{select\_masters\_reduction} refer to the position in this list (starting
with $1$).

An interesting observation is that the runtime to calculate the coefficients of
different master integrals varies strongly.
From experience, the master integrals with the smallest number of lines take
longest.
Therefore, when the back substitution is parallelised across several machines,
those integrals should be evenly distributed among the jobs.
An example to illustrate the usage of the option
\texttt{select\_masters\_reduction} to parallelise the reduction across several
machines is provided in \texttt{examples/topo7\_parallel}. See
\texttt{examples/topo7\_parallel/readme.txt} for details.
When the reduction to all master integrals is complete, the databases can be
merged as described in section \ref{sect:merge}.

\subsection{Command line arguments}

Use \texttt{-{}-parallel=n} where \texttt{n} is at most the number of physical
cores (i.e.\ not counting hyperthreading logical cores) to run \kira{} on
\texttt{n} CPU cores in parallel.

In \kira{}\;\texttt{1.2} option the \texttt{-{}-algebra} has been removed.
The option is now activated automatically where deemed useful as described in
section \ref{sect:defaultoptions}.

\subsection{Integral ordering and basis choice}

Before a large run, try different integral orderings (e.g.\ ordering 1 and 2)
and measure the performance and memory consumption. Note that different
orderings may require different seeds.
In particular, in orderings which produce a ``dot basis'', \texttt{rmax} must be
chosen large enough so that the seed includes the master integrals in each
reduced sector.

Instead or in addition to changing the integral ordering it may also be useful
to choose a different basis of master integrals with the option
\texttt{preferred\_masters}. The choice of the basis can strongly impact the
runtime, because the complexity of the coefficients depends on the basis.

\section*{Acknowledgments}

We thank Peter~Uwer for his continuing support and discussions.
We thank Yang Zhang who inspired the implementation of the algebraic
reconstruction for his support.
We thank Fabian~Lange and Mario~Prausa for their comments on the manuscript and
for testing the new release.
This project has received funding from the European Research Council (ERC) under
the European Union's Horizon 2020 research and innovation programme under grant
agreement No 647356 (CutLoops).
J.U.\ acknowledges support by the ``Phenomenology of Elementary Particle
Phy\-sics beyond the Standard Model'' group at Humboldt-Universit\"at zu Berlin
for providing computing resources.
P.M.\ acknowledges support by the state of Baden-Württemberg through bwHPC and
the German Research Foundation (DFG) through grant no INST 39/963-1 FUGG.
Last but not least we would like to thank all users who helped improving \kira{}
by sending us their comments, feature requests and bug reports.

\appendix

\section{Obtaining Meson and Ninja}
\label{gettingmeson}

If no sufficiently recent version of \texttt{Meson} is available from your
distribution's package repository, the latest version can be installed
system-wide with
\begin{verbatim}
  pip3 install meson
\end{verbatim}
rsp.\ with
\begin{verbatim}
  pip3 install --user meson
\end{verbatim}
in the user's home directory.
\texttt{Python 3.5} or later is required.
The \texttt{Ninja} binary can be downladed from \url{https://ninja-build.org}.

\section{Specifying a compiler and a non-default install location}
\label{compiledependencies}

If you are using a compiler which is not your system compiler, in general you
will also need to compile the dependencies with that compiler.
If the dependencies are not installed in a default location, the search paths
must be set accordingly.

Set the compiler you want to use (here: \texttt{g++-X} for \texttt{C++} and
\texttt{gcc-X} for \texttt{C}) and the installation prefix:
\begin{verbatim}
  export CXX=g++-X
  export CC=gcc-X
  PREFIX=/install/prefix
  export LD_LIBRARY_PATH=$PREFIX/lib64:\
    $PREFIX/lib:$LD_LIBRARY_PATH
  export CPATH=$PREFIX/include:$CPATH
  export PKG_CONFIG_PATH=$PREFIX/lib64/pkgconfig:\
    $PREFIX/lib/pkgconfig:$PKG_CONFIG_PATH
\end{verbatim}
To compile \texttt{CLN} with this compiler and install it to the location
specified by \texttt{\$PREFIX}, extract the \texttt{CLN} source package, change
into the \texttt{CLN} directory and run
\begin{verbatim}
  ./configure --prefix=$PREFIX
  make
  make install
\end{verbatim}

\clearpage

\noindent
Extract the \texttt{GiNaC} source package, change into the \texttt{GiNaC}
directory and run
\begin{verbatim}
  ./configure --prefix=$PREFIX
  make
  make install
\end{verbatim}
Extract the \texttt{yaml-cpp} source package, change into the \texttt{yaml-cpp}
directory and run
\begin{verbatim}
  mkdir build
  cd build
  cmake -DCMAKE_INSTALL_PREFIX=$PREFIX -DBUILD_SHARED_LIBS=ON ..
  make
  make install
\end{verbatim}
Change into the \texttt{\kira{}} directory and run (of course,
\texttt{autotools} may be used as well, see section \ref{installwithautotools})
\begin{verbatim}
  meson --prefix=$PREFIX builddir
  cd builddir
  ninja
  ninja install
\end{verbatim}
To run \texttt{\kira{}}, \texttt{LD\_LIBRARY\_PATH} must be set, so that the same
libraries are linked at startup as were linked at compile time.
Then run
\begin{verbatim}
  $PREFIX/bin/kira
\end{verbatim}
Optionally, you may set the \texttt{PATH} environment variable so that the full
path is not required when starting \kira{}.
\begin{verbatim}
  PATH=$PREFIX/bin:$PATH
  kira
\end{verbatim}
It may be convenient to set and export \texttt{LD\_LIBRARY\_PATH} and
\texttt{PATH} in your shell configuration
(e.g.\ \texttt{\textasciitilde{}/.bashrc} if you are using \texttt{Bash}).


\clearpage


\providecommand{\href}[2]{#2}\begingroup\raggedright\endgroup

\end{document}